# Empathic Chatbot: Emotional Intelligence for Mental Health Well-being

Sarada Devaram
*Faculty of Science & Technology*
*Bournemouth University*
Bournemouth, United Kingdom
s5227932@bournemouth.ac.uk

*Abstract*—Conversational chatbots are Artificial Intelligence (AI)-powered applications that assist users with various tasks by responding in natural language and are prevalent across different industries. Most of the chatbots that we encounter on websites and digital assistants such as Alexa, Siri does not express empathy towards the user, and their ability to empathise remains immature. Lack of empathy towards the user is not critical for a transactional or interactive chatbot, but the bots designed to support mental healthcare patients need to understand the emotional state of the user and tailor the conversations. This research explains the different types of emotional intelligence methodologies adopted in the development of an empathic chatbot and how far they have been adopted and succeeded.

## I. INTRODUCTION

According to the World Health Organization(WHO), 1 in 10 people need mental healthcare worldwide, and different mental disorders are, portrayed by a combination of perceptions, feelings, and relationships with others [1]. Results of a household survey conducted by National Health Services(NHS) states that 1 in 4 people experience a mental health problem each year, 1 in 6 people face a common mental health problem such as anxiety, depression each week in England [6]. The number of people affected is increasing gradually, and with the isolation that Covid-19 brought, the numbers are much higher [7]. Despite having access to health and social services and the number of people who need care is higher, only 70 per 100,000 mental health professionals are available in high-income nations and 2 per 100,000 in low-income nations [5].

The patients distressed with mental conditions struggle to get professional help due to social stigma and hesitation [8]. Furthermore, countries are facing a shortage of mental health professionals [1]. Due to this situation, it is not easy to provide one to one support to treat a patient with a mental health disorder. To overcome these problems, the mental health professionals have adopted the use of technology specifically Artificial Intelligence-based chatbots to address the needs of individuals affected by mental health problems as the first line of defence [2]-[4]. While dealing with a mental health patient, it is vital to understand the emotional state and respond with simple micro-interventions such as suggestions for a deep breathing exercise or a friendly conversation can be useful in increasing the positiveness of patient's mood [9]. The main advantage of these bots is to provide a practical, evidence-based, and an attractive digital solution to help fill the gap of the professional instantly [10]. The evolution of Artificial Intelligence has paved ways for many chatbots, but three therapeutic mental health chatbots [Woebot, Wysa and Tess] are prominent and widely in use [10]. A chatbot programmed to understand emotions might be similarly proactive and keep a history containing that patient's likes and dislikes, or topics that make them laugh and the chatbots could communicate about the likes and dislikes of a patient as situations warrant. Additionally, the adaption of therapeutic chatbots is increasing rapidly due to the following advantages [11]

1. Understand and manage the patient's psychological state and connect them with a health professional during unfavourable events.
2. 24/7 Instant chat support
3. Smart with reactive behaviour such as prompt answering of a question and engage the patients with illness prevention and care tips.
4. Easy to install, configure and maintain and is compatible with various operating systems such as Android, iOS and Linux.
5. For sensitive health care issues, patients might feel less shame and feel more private.
6. Security of personal data is enhanced using different authentication techniques such as login using facial recognition, biometrics or with a passcode.
7. Cost-effective for a few mental conditions, such as stress release.
8. Provide reminders such as taking medication, do exercise, slots for jogging.

## II. TYPES OF EMPATHIC CHATBOTS

A mental health patient can express their feelings using text, emojis or emoticons, voice, recorded audio/video clips or live audio/video. The main aim of the therapeutic chatbots is to understand the appropriate emotions from the user's conversations and suggest them with appropriate treatment or therapy. The empathy expressed by the mental health patient can be cognitive, emotional and compassionate. The purpose of all these categories is to understand the emotions in the user context and relate them to appropriate emotions such as happy, sad, anger, fear [2]. The user emotions can be processed with Artificial Intelligence and deep learning techniques using Natural Language Processing (NLP). NLP depicts how chatbots translate and understand the patient's language. Using NLP, chatbots can make sense of the spoken or written text and accomplish the tasks like keyword extraction, translation, and topic classification. NLP processes the content expressed in natural human language with the help of the techniques such as sentiment analysis, facial recognition and voice recognition [10].

Chatbots with NLP capability can understand the patterns in patient conversation context and analyse the sentiment behind the message by using contextual clues from the voice, video or text input [12].

TABLE I. TYPES OF CHATBOTS

| Methodology | Description |
| --- | --- |
| Sentiment Analysis | Sentiment Analysis extracts opinions, thoughts and emotions from the text or emoticons. |
| Video-based emotion recognition | Facial features extracted from a live or video clip are used to understand the emotions of the patient. |
| Voice-based emotion recognition | Speech features extracted from recorded audio or a phone call are used to understand the emotions of the patient. |

*A. Sentiment Analysis*

Emotion detection is a division of sentiment analysis that deals with the analysis and extraction of emotions. Emotion detection helps mental health professionals to provide tailor-made treatments to their patients. Sentiment analysis recognises how the mental health patient is feeling regarding something. It identifies the patient's message as well as emojis/emoticons as positive, negative or neutral based on the context of the patient's conversation [13]. Extracted opinions are used by the therapeutic chatbots to suggest an appropriate treatment or redirect to a mental health professional in case of any emergencies. The usage of emojis/emoticons increased rapidly in electronic messages [14] and mental health patients can easily express their emotions using smileys and ideograms. An emoticon indicates a deeper meaning in the context of the patient's conversation [14]. The patient responses and emojis/emoticons are converted into a Unicode character set to train the model. The training data collected to train the models vary based on the clinical tools used to gather the data, for instance, collecting the data from clinical records, surveys, and patient's blogs [13]. Depending upon the interpretation of the patient's queries, the categories of the sentiment analysis will be defined, such as aspect-based sentiment analysis is used to analyse the text-based messages from the patient response and emotion-based analysis is used to analyse emoticons [15].

Advantages

- Chatbots provide treatment analysis for a patient by analysing their responses, whether the treatment is causing negative or positive effects on the patient.
- Chatbots will have a clear overview of the patient's emotional state and respond accordingly.
- Chatbots identify what messages and conversations act as emotive triggers that change the patient's mood.
- Chatbots can identify emergencies such as suicidal thoughts and escalate or redirect them to appropriate professionals.

Limitations

- Multiple sentiments in one sentence is complicated for sentiment annotations [15].

- Defining neutral: Sometimes, the patient does not show any indication of their emotional state, but they describe situations then the chatbot will have the difficulties to consider that the patient is in a negative emotional state.

*B. Video-based emotion analysis using facial recognition*

In the process of patient-therapeutic chatbot video-based interactions, it is vital to detect, process and analyse the patient's emotions and perceptions to adjust the treatment strategies. The goal of facial recognition is to collect data and analyse the feelings of patients to make relevant responses possible. The data is gathered from various physical features such as body movements, facial expressions, eye contact and other physical, biological signals. These physical emotions are classified into different categories, such as sadness, happiness, surprise, fear, and anger. Image processing and computer vision techniques are used to extract the mental health patient's facial features using two approaches – Geometric-based, appearance-based [16].

The geometric-based approach signifies the mental health patient face's geometry by extracting the nodal points, the shapes and the positions of the facial components like eyebrows, eyes, mouth, cheeks and nose then compute the total distance among facial components to create an input feature vector. The main challenge with this approach is to gain high accuracy in facial component detection in real-time.

The appearance-based approach indicates mental health patient face textures by extracting the variations in skin textures and face appearances. This approach uses Local Binary Patterns (LBP), Local Directional Patterns (LDP), and Directional Ternary Patterns (DTP) to encode the textures as training data. The empathic chatbot detects the emotions from facial expressions using appearance-based approach because a geometry-based approach needs reliable and accurate facial component detection to gain maximum accuracy value which is very difficult in real-time scenarios [17].

The process of developing a video-based emotional system is as follows [16]

1. The chatbot detects the patient's face from the video chat.
2. The facial appearance detection gets the patient's facial features and converts them to input feature vectors.
3. The selected Machine Learning(ML) classifier categorises the patient emotions into different classes such as sadness, happiness, disgust, neutral, anger, surprise and fear.
4. Finally, the accuracy metrics are calculated for subsequent analysis.

Advantages

- It provides the flexibility of the appointments by reducing the physical contact, or direct physician interaction.
- It makes it easier to organise appointments.

- It improves medical treatment by examining subtle facial traits, facial recognition.

Limitations

- The accuracy may vary when a patient changes appearance, or the camera angle is not quite right.

*C. Voice-based emotion identification*

The empathic chatbots should understand the emotions from the context of the patient's voice calls or recorded audio files. The chatbot is equipped to access the sensor/microphone, which can capture the voice sample on the patient's behaviour without having to interpret the inputs. The emotions are identified using two classes of speech features such as the lexical and acoustic features [16].

- Lexical speech features: These features are bound with the vocabulary used by the mental health patient. Lexical features need the text extraction from the speech to predict the patient's emotions so it can be used on the recorded audio files.

- Acoustic speech features: These features are bound with the pitch, jitter and tone of the mental health patient. Acoustic features need the audio data for understanding the emotions in the patient's conversation so it can be used for voice calls with the patient. The acoustic model will be trained to extract the spectral features from speech signals.

A voice-based emotion recognition system is a pattern recognition system which consists of three principal parts: processing the audio signals, feature calculation and voice classification. The main aim of signal processing includes the digitisation of the audio signal, filtering the audio signal and the segmentation of the audio conversation of spoken words into text. The feature calculation aims to find the properties of the pre-processed and digitised acoustic signal that represent the emotions and convert them into an encoded vector. Finally, the Machine Learning(ML) classification algorithms will be used on the feature selection vector. These classification algorithms can vary based on the trained dataset [18].

Advantages

- Chatbots can often detect a mental health patient emotion even if it cannot understand the language because the acoustic speech features use voice elements such as pitch and tone.

- Patients find it works faster than typing the text messages.

Limitations

- Sometimes it is difficult to analyse the speech elements such a patient can express anger in slow pitch and tone, which makes it challenging to identify the emotion behind the speech.

III. THE SUCCESS OF THE EMPATHIC CHATBOTS

The proliferation of chatbots that are dedicated to helping mentally and emotionally distressed people indicates that seeking help online is becoming increasingly popular. The empathic chatbots developed on Cognitive Behavioural Therapy (CBT) platform, which essentially means therapy through conversation. The therapy aims to turn the patient's negative thoughts into positive ones, by initiating a joyful daily talk and creating a relaxing environment for the patient. Patients with emotional distress are more comfortable talking anonymously to a machine from the comfort of their home without the fear of being judged, than physically visiting a psychologist's office, which is already stigmatised in many societies around the world [19].

WoeBot, Wysa and Tess are few prominent chatbots that are helping the anxiety and depression patients [10].

WoeBot is an AI application that claims to help alleviate mental health disorders through fully automated conversations. The application's conversational agent initiates the chat by asking users how there are feeling and sends them tips and videos on wellbeing according to their needs. Surveys conducted on Woebot users by Stanford University indicate a significant improvement with feelings of depression and anxiety [20].

Wysa is an AI-powered bot that helps users manage their feelings through Cognitive Behavioural Therapy (CBT), Dialectical Behavior Therapy (DBT), and simple exercises [21].

Tess is a psychological AI-powered chatbot that focuses on mental health and emotional wellness. Tess does not work on the pre-programmed responses; it understands the situation and response according to user preferences, and also it remembers user's likes, dislikes and poses an understanding attitude [22].

There are many more apps that claim to cater to one's wellbeing through conversing and analysing mood, physical activities, movement patterns, energy, social interactions and locations [19].

IV. LIMITATIONS OF THE EMPATHIC CHATBOTS

- One of the main challenges being faced are the contextual awareness during the patient conversations; lack of contextual data for training, changes in patient's conversational behaviour with emojis, short descriptions/abbreviated texts during the discussions.

- Few healthcare professionals argue that AI should be supplemented instead of replacing the health professionals and finding an appropriate role of AI is a significant challenge for the future. [23]

- Limited Adoption – Many health professionals in the US indicated that the bots cannot effectively understand the needs of patients and cannot be responsible for a thorough diagnosis. Some think that the usage of chatbots in health care might pose a risk of self-diagnosis and failing to understand the diagnosis [24]

- Other challenges are confidentiality and patient privacy. Since patient conversation includes personal matters, it is essential to encrypt patient conversations or anonymise the patient data in the database.

## V. Conclusion

While an empathic chatbot may offer a mental health patient with a forum to discuss problems and provide access to help guides and also increase mental health literacy and a way to track moods, but an empathic chatbot is not an alternative of a mental health professional or a therapist. Despite the few limitations, empathic chatbots are proving a nascent technology with massive future potential. The empathy added to a chatbot has filled a clear and critical gap that is already proving life-changing for patients. These chatbots show how it is possible to leverage conversational AI in different ways. So, empathy chatbots in mental health are not just making waves in the healthcare industry, but they are also paving the way for more innovative and beneficial uses of chatbot technology in all aspects of life.

This research explained the various methods of AI techniques which are applied to classify the intent of the conversation into emotions and explored a few prominent apps in this space.